\documentclass[amsmath,amssymb,twocolumn,floatfix]{revtex4-2}

\usepackage{graphicx}
\usepackage{dcolumn}
\usepackage{bm}
\usepackage{hyperref}

\providecommand{\dif}[0]{\mathrm{d}}

\providecommand{\order}[1]{{\cal O} \left( #1 \right)}

\providecommand{\pgas}[0]{P_{\text{gas}}}
\providecommand{\rhogas}[0]{\rho_{\text{gas}}}
\providecommand{\rhogas}[0]{\rho_{\text{gas}}}

\providecommand{\rvir}[0]{R_{\text{vir}}}
\providecommand{\tvir}[0]{T_{\text{vir}}}
\providecommand{\mvir}[0]{M_{\text{vir}}}
\providecommand{\vvir}[0]{V_{\text{vir}}}

\providecommand{\mpr}[0]{m_{\rm p}}

\providecommand{\nel}[0]{n_{\rm e}}
\providecommand{\np}[0]{n_{\rm p}}
\providecommand{\nh}[0]{n_{\rm He}}
\providecommand{\nj}[0]{n_j}

\providecommand{\ngas}[0]{n_{\rm gas}}

\providecommand{\pi}[0]{p_i}

\providecommand{\tcool}[0]{t_{\rm cool}}
\providecommand{\rcool}[0]{r_{\rm cool}}
\providecommand{\mcool}[0]{M_{\rm cool}}
\providecommand{\mcold}[0]{M_{\rm cold}}
\providecommand{\mstar}[0]{M_{\rm star}}
\providecommand{\tdyn}[0]{t_{\rm dyn}}
\providecommand{\mhalo}[0]{M_{\rm halo}}
\providecommand{\mhalotod}[0]{M_{{\rm halo},0}}

\providecommand{\msun}[0]{M_{\rm sun}}
\providecommand{\lambdanet}[0]{\Lambda_{\rm net}}
\providecommand{\zgas}[0]{Z_{\rm gas}}
\providecommand{\tav}[0]{t_{\rm avail}}

\begin{document}

\title{Effects of helium sedimentation on late star formation in galaxy clusters}

\author{J.~Racker}
\email{jracker@unc.edu.ar}
\affiliation{Universidad Nacional de C\'ordoba (UNC). Observatorio Astron\'omico de C\'ordoba (OAC). C\'ordoba, Argentina
}

\affiliation{Consejo Nacional de Investigaciones Cient\'{\i}ficas y T\'ecnicas (CONICET), Instituto de Astronom\'{\i}a Te\'orica y Experimental (IATE), Laprida 854, X5000BGR, Córdoba, Argentina
}
\author{N.~Padilla}
\email{nelson.padilla@unc.edu.ar}
\affiliation{Universidad Nacional de C\'ordoba (UNC). Observatorio Astron\'omico de C\'ordoba (OAC). C\'ordoba, Argentina
}

\affiliation{Consejo Nacional de Investigaciones Cient\'{\i}ficas y T\'ecnicas (CONICET), Instituto de Astronom\'{\i}a Te\'orica y Experimental (IATE), Laprida 854, X5000BGR, Córdoba, Argentina
}

\begin{abstract}
We discuss how helium sedimentation in galaxy clusters can affect the history of star formation in the central cluster galaxy. As helium sediments, the gas density in the inner regions of the cluster increases and there is also a non-trivial, radially dependent redistribution of the atomic nuclei and electrons. As a result, the cooling rate in the center increases and this can enhance star formation. 
On the other hand, there is a slow contraction of the intracluster gas, which may induce  
gravitational heating  and therefore has an opposite effect on star formation. In this work we present these effects and aim to estimate their relevance. 
For this we have performed a 1-dimensional numerical simulation of helium sedimentation and applied it to a simple semi-analytical model of star formation. We find that for clusters with a halo mass $\mhalo \lesssim 10^{14} \msun$, helium sedimentation effects on the star formation rate are negligible, even under idealized conditions. In the intermediate range, $10^{14} \msun \lesssim \mhalo \lesssim 10^{15} \msun$, the effects are at most mild, below a factor $\sim 2$ in the isothermal model we consider, even for idealized conditions. For clusters with a halo mass $\mhalo \gtrsim 10^{15} \msun$, helium sedimentation effects can potentially be very important and renew star formation activity in the central  galaxy. 
\end{abstract}

\maketitle

\section{Introduction}
\label{sec:introduction}

The brightest cluster galaxy (BCG) grows at the center of a hot, stratified atmosphere, where cooling, feedback, and gravitational settling act together. In semi-analytic models (SAMs), the build-up of massive centrals follows models for gas cooling, mergers, and active galactic nucleus (AGN) regulation which are calibrated using observational data on galaxies and clusters 
(e.g. see \citealt{Cole2000,Springel2001,Croton2006,Bower2006,Somerville2008,Lacey2016}). Recent developments include molecular-gas based star-formation laws and prescriptions tuned for massive centrals (\citealt{DeLucia2007,Lagos2011a,Lagos2011b}). None of these schemes tracks helium sedimentation in the intracluster medium (ICM).

Cooling cores provide the broader context. Short central cooling times are balanced, on average, by intermittent AGN heating, which limits classical cooling flows and keeps most BCGs quiescent (as reviewed by \citet{Fabian1994} and \citet{McNamara2007}). Residual star formation is, however, observed in low-entropy cores and correlates with cooling-time thresholds (e.g.~\citealt{Rafferty2008,ODea2008,VOit2008}). Small changes to density or mean molecular weight near the center can therefore move systems across condensation criteria.

In a multicomponent, proton-dominated plasma, helium tends to drift inward under gravity.
 As a result, the gas becomes inhomogeneous and the density increases in the inner regions. 
Some of the concomitant effects have already been explored in the literature, including biases in X-ray and Sunyaev–Zel'dovich inferences and modified thermodynamics, with the magnitude set by transport physics and suppression by turbulence or magnetic fields (see \citealt{ChuzhoyNusser2003,ChuzhoyLoeb2004,EttoriFabian2006,Peng2009,Shtykovskiy2010,Bulbul2011,Medvedev2014}). 
However, the effects of helium sedimentation  on the star formation history of BCGs have not been explored so far. In particular,  they have not been included in SAMs or in hydrodynamical simulations. In this work we identify two competing tendencies that result from helium sedimentation. On one hand, the enhanced central gas density shortens cooling times (at fixed temperature). On the other hand,  the slow contraction linked to sedimentation releases gravitational potential energy that can heat the gas.

These complex processes merit a focused study of helium sedimentation and its effect on the star formation activity of BCGs.
In order to obtain quantitative estimates,
we simulate helium sedimentation in cluster atmospheres and couple the resulting mean molecular weight and density evolution to a simple semi-analytic star-formation model for massive centrals. Our goal is to quantify how BCG fueling histories change when helium is allowed to drift, relative to otherwise identical models with fixed primordial composition. In this work we deliberately switch off AGN feedback to isolate the impact of helium sedimentation on cooling and BCG growth; a more realistic treatment is deferred. 

The remainder of this paper is organized as follows. In Sec.~\ref{sec:helium_sedimentation} we outline the equations for helium sedimentation and the numerical simulation we have performed to solve them. In Sec.~\ref{sec:model_star_formation} we set the SAM we use to estimate the stellar mass evolution. Next, in Sec.~\ref{sec:helium_and_star_formation}, we describe the main results. Finally, the conclusions and an outlook are given in Sec.~\ref{sec:conclusions}.


\section{Helium sedimentation}
\label{sec:helium_sedimentation}
The ICM is mainly composed of hydrogen and helium. The helium, being heavier than hydrogen, tends to concentrate more in the inner regions, however this sedimentation process in galaxy clusters is very slow, with time-scales of the order of the age of the Universe or larger. The diffusion of particles in a cluster can be described by Burgers equations. Assuming spherical symmetry, each species $s$ satisfies an equation of continuity and momentum conservation (see e.g.~\citet{Peng2009}):
\begin{eqnarray}
    \frac{\partial n_s}{\partial t} +  \frac{1}{r^2} \frac{\partial \left(r^2 n_s u_s\right)}{\partial r}&=& 0 \, , \label{eq:continuity} \\
    \frac{\partial P_s}{\partial r} + n_s A_s m_p g - n_s Z_s e E &=& \sum_t K_{st} (w_t-w_s), \label{eq:momentum_conservation}
\end{eqnarray}
where $r$ is the radial coordinate and $t$ the time. For each species $s$ we have denoted its mass by $A_s\, m_p$ (with $m_p$ the proton mass), its charge by $Z_s\, e$,  its number density by $n_s$, its pressure by $P_s$, and its velocity by $u_s$. The acceleration due to the gravitational field is represented by $g$ and the electric field by $E$ (the existence of this global electric field in plasmas subject to a gravitational field was first studied in~\citealt{Pannekoek22, Rosseland24}, and~\citet{Eddington1926}). The diffusion velocity $w_s$ is defined as the velocity of species $s$ relative to the center-of-mass velocity of the composite fluid ($u$), i.e.,
\begin{equation}
    w_s = u_s - u \, ,
\end{equation}
where
\begin{equation}
    u = \frac{\sum_s n_s \, A_s \, u_s}{\sum_s n_s \, A_s} \, .
\end{equation}
The diffusion velocities satisfy the following relations:
\begin{eqnarray}
    \sum_s A_s \, n_s \, w_s & = & 0\, , \label{eq:mass_conservation}\\
    \sum_s Z_s \, n_s \, w_s & = & 0\, , \label{eq:charge_conservation}
\end{eqnarray}
where the first one comes from the definitions and the second one from the quasi-neutrality condition. Finally, the resistance coefficient, $K_{st}$, in the absence of magnetic fields and turbulence is given by
\begin{equation}
\label{eq:resistance_coefficient}
    K_{st} \simeq \frac{ 4 \sqrt{2 \pi}}{3} \, \frac{Z_s^2 Z_t^2 e^4 \mu_{st}^{1/2}}{\left( k_B T \right)^{3/2}} \, n_s \, n_t \ln \Lambda_{st} \, ,
\end{equation}
where $\mu_{st}= A_s A_t m_p /(A_s+A_t)$ is the corresponding reduced mass. Moreover, $\Lambda_{st}$ is the Coulomb logarithm, which has a mild dependence on temperature, density and composition. For the ICM conditions we consider, a typical value is $\Lambda_{st} \sim 40$. 

From Eqs.~\eqref{eq:mass_conservation} and~\eqref{eq:charge_conservation} it is clear to see that, in the center-of-mass frame of a fluid element, for every helium nucleus falling to the center, there are roughly 4 protons and two electrons going up. This radial  redistribution in the number of particles causes a transient change in the pressure gradient and consequently a break of hydrostatic equilibrium which, however, tends to recover on a time-scale much shorter than that of the diffusion process. This results in a slow contraction of the intracluster gas, given by
\begin{equation}
\label{eq:ucm_rate_of_change}
    \frac{\dif u}{\dif t} = - \frac{1}{\rhogas} \, \frac{\partial \pgas}{\partial r} - g \;,
\end{equation}
where $\pgas$ and $\rhogas$ are the total pressure and mass density of the gas, i.e., 
\begin{eqnarray}
    \pgas &=& \sum_s P_s = \sum_s n_s k_B T, \\
    \rhogas &=& \sum_s \rho_s =  \sum_s A_s m_p n_s,
\end{eqnarray}
with $T$ denoting the temperature.

Following these equations, we have performed a 1-dimensional numerical simulation of helium sedimentation for a completely ionized gas composed of protons (p), helium (He), and electrons (e), under isothermal conditions. The gravitational potential is obtained from a Navarro–Frenk–White profile of dark matter distribution (\citet{nfw96}), with the concentration parameter obtained from the analytical formula in Appendix C of~\citet{Ludlow2016}. The contribution of the gas to the gravitational potential is neglected. For a given dark matter halo mass we calculate the virial radius (according to~\citet{Bryan1998}), $\rvir$, and define a radial grid going from $10^{-5}\, \rvir$ to $\rvir$, divided in around 1000 cells.  

At the initial simulation time, $t=t_i$, the gas is taken to be in hydrostatic equilibrium with a homogeneous and primordial chemical composition. Therefore, the initial gas density, $\ngas=\sum_s n_s$, satisfies
\begin{equation}
\label{eq:heg}
    \frac{\partial \pgas}{\partial r} = - \mu \, \mpr \ngas \frac{\partial \phi}{\partial r},
\end{equation}
where $\phi$ is the gravitational potential
and the mean molecular weight of the gas, $\mu$, is given by
\begin{equation}
\label{eq:mu}
    \mu = \mu(r) \equiv \frac{A_e \nel + \sum_j A_j n_j}{\ngas} \approx \frac{\sum_j A_j n_j}{\ngas}.
\end{equation}
The solution to Eq.~\eqref{eq:heg} with constant temperature and molecular weight, $\mu=\mu_i$, sets our initial conditions, namely,
\begin{equation}
\label{eq:initial_gas_density}
\ngas(r,t_i) = \ngas^0 \, e^{-\mu_i \mpr \phi(r)/(k_B T)}\, .  
\end{equation} 
The constant $n_0$ is determined by the condition that
\begin{equation}
\label{eq:n0}
    \int_0^{\rvir} \ngas(r,t_i)\, 4 \pi r^2 \, \dif r = \frac{\Omega_b}{\Omega_{\rm DM}} \, \frac{\mvir}{ \mu_i \, \mpr},
\end{equation}
where $\mvir$ is the virial mass.
In turn, the initial proton, helium, and electron densities are just proportional to the gas density, namely,
\begin{eqnarray}
    \np (r,t_i) &=& f_{\rm p} \, \ngas(r,t_i), \\
    \nh (r,t_i) &=& f_{\rm He} \, \ngas(r,t_i), \\
    \nel (r,t_i) &=& \np (r,t_i) + 2 \,\nh(r,t_i),
\end{eqnarray}
where the last equality comes from the requirement of local quasi-neutrality.
In this work we have taken $\Omega_b\, h^2= 0.0224$, $\Omega_{\rm DM} h^2 = 0.12 $, $h=0.674$ (from~\citet{Planck2018VI}), and $\mu_i=0.59$, which corresponds to $f_{\rm He}/f_{\rm p} \simeq 3/38$. 

Then, at each time step and for given values of the densities $n_s(r)$  ($s=$ e, p, He), we obtain the diffusion velocities from Eqs.~\eqref{eq:momentum_conservation}, \eqref{eq:mass_conservation}, and~\eqref{eq:charge_conservation} neglecting the mass of the electron, so that, approximately, 
\begin{eqnarray}
    E &=& \frac{1}{Z_e e \, \nel} \frac{\partial P_e}{\partial r}\, ,\\
    w_{\rm He} &=& -\frac{K_{{\rm p} {\rm He}}^{-1}}{\left(1+\frac{A_{\rm He} \, \nh}{A_{\rm p} \, \np}\right)} \Biggl[ \frac{\partial P_{\rm He}}{\partial r} + \nh \, A_{\rm He} \, \mpr\, g \Biggr.\notag \\ \Biggl. &\quad&  - \nh \, Z_{\rm He} \, e \, E \Biggr], \\
    w_{\rm p} &=& - \,\frac{A_{\rm He} \, \nh}{A_{\rm p} \, \np} \, w_{\rm He} \,.
\end{eqnarray}
In turn, the center-of-mass velocity, $u(r)$, is calculated from Eq.~\eqref{eq:ucm_rate_of_change}. Finally, at each time step the number densities of protons and helium are updated from the continuity Eqs.~\eqref{eq:continuity}, which we solve numerically simulating an actual flux of particles. The electron density is always established by the quasi-neutrality condition, namely, $\nel=\np + 2 \nh$. The only border conditions we impose are zero flux of proton and helium particles at the borders, i.e., at $r= 10^{-5} \rvir$ and $r=\rvir$. 


\section{Semi-analytical model for star formation}
\label{sec:model_star_formation}

In this section we introduce a simplified version of a semi-analytic model  designed to isolate the impact of helium sedimentation on the star-formation histories of massive cluster galaxies. 
We explicitly track the main components of the ICM (protons and helium nuclei), while keeping all other prescriptions identical in a reference model that includes no sedimentation. To highlight the maximal effect of the modified gas density and chemical composition, we intentionally do not include AGN feedback, so that cooling flows and star formation respond only to the change in gas thermodynamics and mean molecular weight at fixed cosmology \citep{Planck2018VI}. 
In what follows we give details on our simplified model which is adapted from~\citet{Springel2001}.

Our model is based on simple definitions, starting with the cooling time, which is a measure of the time scale required for the gas at each radius to cool due to radiative emission. It is given by 
\begin{equation*}
    \tcool (r) = \frac{3 \ngas k_B T/2}{\lambdanet (n_s, T, \zgas)},
\end{equation*}
where the cooling function, $\lambdanet$, is the energy radiated per unit time and volume. In this work we use the standard collisional ionization  equilibrium cooling function given in the table 6 of~\citet{Dopita1993}, for zero metals, but with a correction factor to take into account the inhomogeneous gas composition that results as helium sediments. This factor can be approximated by noticing that helium sedimentation can only be significant for the most massive galaxy clusters, for which the intracluster gas is completely ionized and the X-ray emission is dominated by thermal bremsstrahlung. In this case (and neglecting the dependence of the Gaunt factor on the temperature and charges of the ions), the total energy emitted per unit time and volume is proportional to
\begin{equation}
\lambdanet \propto \sum_j Z_j^2 \, \nel \, \nj,
\end{equation}
where the summation goes over all ion species, i.e., protons and helium nuclei in our case.  
Then, we  calculate the cooling function taking the values tabulated in~\citet{Dopita1993}, dividing by the sum $\sum_{j={\rm p, He}} Z_j^2 \, \nel \, \nj$ computed using the densities also given in table 6 of~\cite{Dopita1993}, and multiplying by the sum evaluated using the densities obtained from our simulation.   

Based on the cooling time it is possible to define the cooling radius, $\rcool$, as the radius at which the cooling time is equal to a typical time scale of the cluster, $\tav$, which we take here as the age of the cluster for most of the analysis. Therefore the cooling radius is defined through the equation
\begin{equation}
    \tcool\, (\rcool) = \tav = t\, .
\end{equation}
In case $\tcool(r) < t$ for all $r \le \rvir$, we take $\rcool = \rvir$, while if $\tcool(r) > t$ for all $r \le \rvir$, then $\rcool = 0$.
Furthermore, we denote by $\mcool$ the total mass of the gas that (roughly) has had enough time to cool, which is the total mass of gas inside a sphere of radius $\rcool$, i.e., 
\begin{equation}
    \mcool = \int_0^{\rcool} \rhogas (r) \, 4\pi r^2 \, \dif r.
\end{equation}
All the quantities we have defined actually depend on time because we consider clusters with a certain history of mass growth (to be described below), and also because the gas density profile changes due to helium sedimentation.  

The star formation rate in a galaxy depends on the amount of cold gas and can be modeled as~\citep{Springel2001}
\begin{equation}
\label{eq:mstar_rate}
    \frac{\dif \mstar}{\dif t} = \alpha \frac{\mcold}{\tdyn}\,,
\end{equation}
where $\mstar$ is the total mass in stars (at a given time), $\alpha$ is a parameter related to  the efficiency of star formation, and $\mcold$ is the total amount of cold gas. The dynamical time of the BCG, $\tdyn$, can be approximated roughly by $\tdyn = 0.1 \rvir/\vvir$, where $\vvir$ is the virial velocity.
With no AGN accretion and feedback, $\mcold$ is just that part of $\mcool$ that has not been converted into stars and therefore its rate of change is given by
\begin{equation}
\label{eq:mcold_rate}
    \frac{\dif \mcold}{\dif t} =  \frac{\dif \mcool}{\dif t} - \alpha \frac{\mcold}{\tdyn}\, .
\end{equation}
We have solved the Eqs.~\eqref{eq:mstar_rate} and~\eqref{eq:mcold_rate} numerically, with the initial conditions $\mstar(t_i) = \mcold(t_i) =0$.

Finally, we specify that in our simplified model galaxy clusters grow according to the mean mass growth rate given in~\citet{fakhouri2010}, namely
\begin{eqnarray}
<\dot \mhalo> & = & 46.1 \, \msun \; {\rm yr}^{-1} \left( \frac{\mhalo}{10^{12} \msun} \right)^{1.1}  \notag \\
&\times& (1+1.11 z) \, \sqrt{\Omega_{m}(1+z)^3 + \Omega_{\Lambda}}\, .
\end{eqnarray}
Using that $\frac{\dif \mhalo}{\dif z} =  \frac{\dif \mhalo}{\dif t} \frac{\dif t}{\dif z}= - <\dot \mhalo> [H(z) (1+z)]^{-1}$, we can integrate this equation to obtain the evolution of the mass of a dark matter halo having a present-day mass equal to $\mhalotod$: 
\begin{eqnarray}
    \mhalo (z) &=& \Biggl[  \mhalotod^{-0.1} + \frac{4.61 \, {\rm yr}^{-1}}{10^{13.2}\,\msun^{0.1}\, H_0} \Biggr. \notag \\ \Biggl.
    &\times&  (1.11 z - 0.11 \ln (1+z)) \Biggr]^{-10} \;. 
    \label{Eq:Mhdot}
\end{eqnarray}
Even though this choice includes no mergers as using merger trees would, it is in line with our target of simplicity.  We incorporate a sedimentation suppression factor $f_B$ in the following section which can be loosely interpreted as including merger activity, to some degree: it is thought that clusters of galaxies that have undergone a recent merger are less relaxed, have higher turbulence, and this suppresses the ability of gas to sediment.  Adding the suppression factor as a free parameter of our model allows us to include indirectly the effects of mergers on sedimentation. 


\section{Results}
\label{sec:helium_and_star_formation}

Helium sedimentation changes the density profile of the intracluster gas, increasing the density -and henceforth the cooling rate- in the central regions, which can induce an increased rate of star formation, particularly at late times, when helium might have had enough time to sediment, at least partially. 
A more subtle effect is that there is a loss of gravitational potential energy as the intracluster gas slowly contracts, which can result in a source of heating, partially inhibiting star formation. 

To perform a quantitative estimate of these effects, we have used the SAM presented in the previous section, considering at each time step two different gas density profiles to determine the cooling radius and related quantities. One of them corresponds to the absence of helium sedimentation and it is therefore chemically homogeneous, having a radial-independent molecular weight. For an isothermal ideal gas, the gas density at time $t$ is given by Eqs.~\eqref{eq:initial_gas_density} and~\eqref{eq:n0}, where $\mvir=\mvir (t)$ and $\rvir = \rvir (t)$ are the virial mass and radius at time $t$ of a galaxy cluster that evolves according to Eq. \ref{Eq:Mhdot} and that at present times ($t=t_0, \, z=0$) has a given halo mass $\mhalotod$. Moreover, in most of the analysis we have taken the temperature equal to the virial temperature, $T(t)=\tvir (t)$. The other density profile is obtained, at each time $t$, considering again a galaxy cluster with $\mvir=\mvir (t)$ and $\rvir = \rvir (t)$, but where helium sedimentation took place during a time period equal to $t$ (which is taken to be the age of the cluster at time $t$). To obtain this sedimented density profile we perform, at each time step, the complete numerical simulation described in Sec.~\ref{sec:helium_sedimentation}, 
taking as the initial density profile, the homogeneous one that we had just described, i.e., the one obtained for $\mvir(t), \rvir(t)$ and $\tvir(t)$.
We note that a more realistic simulation of a cluster growth and helium sedimentation is expected to yield a more homogeneous gas distribution, i.e., with less sedimentation, since halos grow via accretion of external gas and  mergers with smaller halos. As mentioned above, we have included this indirectly by introducing a suppression of the sedimentation.  The case of no suppression can be taken as an upper bound of the effect of sedimentation.

To estimate the effect of the gravitational heating associated to helium sedimentation, we have also solved the equations of the SAM, but adding a heating term in the Eq.~\eqref{eq:mcold_rate}, as commonly done to model AGN feedback \citep{Croton2006}, namely 
\begin{equation}
\label{eq:mcold_rate_with_heating}
    \frac{\dif \mcold}{\dif t}\Bigr|_{h} =  \frac{\dif \mcool}{\dif t} - \alpha \frac{\mcold}{\tdyn} - \frac{\dot E_{hs}}{1.5\, k_B T/(\mu \, \mpr)}\, .
\end{equation}
Here $1.5\, k_B T/(\mu \, \mpr)$ is the thermal energy per unit mass and $\dot E_{hs}$ is the heating power, which we take equal to the rate of change of the gravitational potential energy of the gas inside a sphere of radius $r = \rcool (t)$. This corresponds to the assumption that all the gravitational energy that is lost by the gas inside $\sim \rcool$ contributes to heating the gas. 

\begin{figure}[t]
\includegraphics[width=0.5\textwidth,{angle=0}]{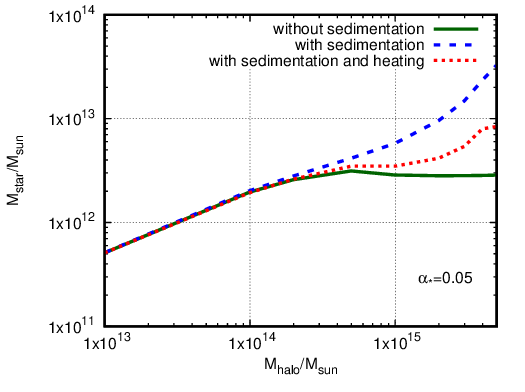}
\caption{\label{fig:stellar_halo_mass_relation} Stellar mass as a function of halo mass at redshift $z=0$, both in solar mass units. The stellar-halo mass relation without helium sedimentation, with helium sedimentation but without heating, and with helium sedimentation and heating are given by the green solid, blue dashed, and red dotted lines, respectively. Here we have taken $\alpha = 0.05$.}
\end{figure}
\begin{figure*}[t]
\includegraphics[width=0.45\textwidth]{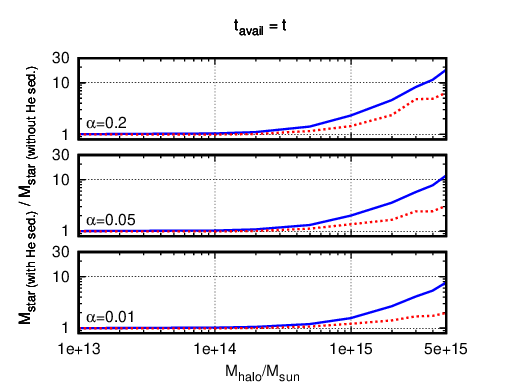} \quad
\includegraphics[width=0.45\textwidth,{angle=0}]{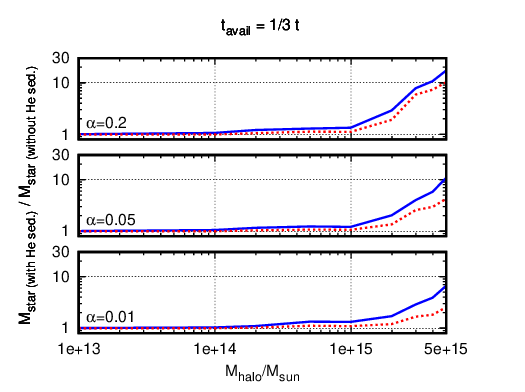}
\caption{\label{fig:stellar_halo_mass_relation2} Ratio between stellar mass with helium sedimentation and stellar mass without helium sedimentation as a function of halo mass, at redshift $z=0$, without including heating (blue solid curves) and including heating (red dotted curves). Each panel corresponds to a different value of $\alpha$, $\alpha=0.01, 0.05$, and $0.2$, from bottom to top. The left plot is for $\tav=t$ and the right plot for $\tav=t/3$.}
\end{figure*}
 
In Fig.~\ref{fig:stellar_halo_mass_relation} we show the resultant stellar-halo mass relation  without helium sedimentation, with helium sedimentation and no heating, and with helium sedimentation including  the  associated gravitational heating. Notice that the stellar masses are somewhat larger than in models with AGN feedback, as expected since the effect of this type of feedback is to decrease the ability of the BCG to increase its stellar mass (see e.g.~\citet{Bower2006}); this figure is solely intended to isolate the possible effects of helium sedimentation. 
Fig.~\ref{fig:stellar_halo_mass_relation2}  shows the ratio between stellar mass with and without helium sedimentation, 
for different values of the the star-formation efficiency parameter $\alpha$ within a factor of a few from the fiducial value $\alpha=0.05$. The two plots in this figure correspond to two different choices of the time scale used for defining the cooling radius, $\tav=t$ and $\tav= 1/3 \,t$, to cover a range of values, as this choice is not unique in the literature and it can be taken as another parameter of the model.

Some conclusions can be inferred from these figures: (I)~Given the procedure we have implemented for comparing star formation with and without helium sedimentation and the majority of the assumptions we have made (except for constant temperature, see below), it seems more appropriate to interpret the results as conservative upper bounds on helium sedimentation effects. Therefore Figs.~\ref{fig:stellar_halo_mass_relation} and~\ref{fig:stellar_halo_mass_relation2} show that helium sedimentation effects on star formation are negligible for galaxy clusters with $\mhalo \lesssim 10^{14} \msun$. 
(II)~For clusters with masses in the range $10^{14} \msun \lesssim \mhalo \lesssim 10^{15} \msun$ the effects are moderate or absent, although it could be interesting to consider more realistic temperature profiles (e.g.~\citet{nelson}) to draw more definite conclusions. 
(III)~For the most massive clusters, $\mhalo \gtrsim 10^{15} \msun$, the effects can potentially be very large and even $\order{10}$ effects seem in principle plausible.
(IV)~When helium sedimentation effects are significant, the associated gravitational heating can be relevant and should be included in order to obtain the net effect on star formation coming from the sedimentation process. This also motivates to develop a more detailed treatment of this heating effect. Note in particular the dependence on $\tav$ that is apparent from Fig.~\ref{fig:stellar_halo_mass_relation2}: a smaller $\tav$ implies a smaller $\rcool$ and therefore a reduced heating when estimated as described after Eq.~\eqref{eq:mcold_rate_with_heating}.

\begin{figure*}[t]
\includegraphics[width=0.45\textwidth]{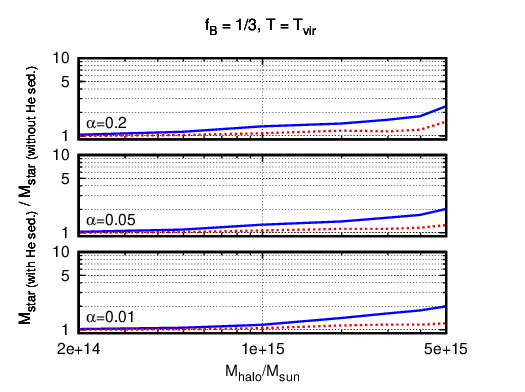} \quad
\includegraphics[width=0.45\textwidth,{angle=0}]{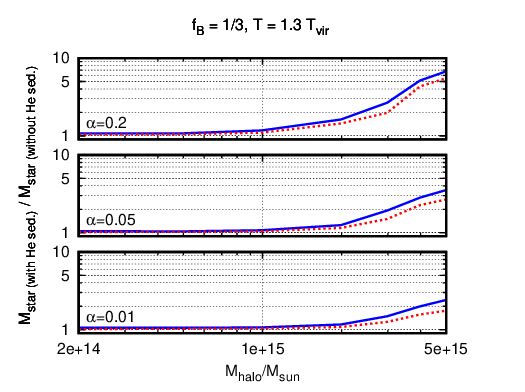}
\caption{\label{fig:stellar_halo_mass_relation3} Ratio between stellar mass with helium sedimentation and stellar mass without helium sedimentation as a function of halo mass, for a suppression factor $f_B=1/3$, at redshift $z=0$, without including heating (blue solid curves) and including heating (red dotted curves). Each panel corresponds to a different value of $\alpha$, $\alpha=0.01, 0.05$, and $0.2$, from bottom to top. The left plot is for $T=\tvir$ and the right plot for $T=1.3 \, \tvir$.}
\end{figure*}

In Figs.~\ref{fig:stellar_halo_mass_relation} and~\ref{fig:stellar_halo_mass_relation2} we have assumed that the resistance coefficient $K_{\rm{p He}}$ is given by Eq.~\eqref{eq:resistance_coefficient}, with no suppression of transport by magnetic fields and plasma instabilities (see e.g.~\citealt{Berlok2015, Berlok2016}).
Some observations suggest that transport coefficients, like the viscosity, might be very suppressed in the ICM, below 25\% (\citet{Ignesti24}) or by more than a factor 10 in the Coma cluster (\citet{Zhuravleva19}), with respect to the idealized values (see also the discussions in, e.g., \citealt{ChuzhoyLoeb2004, Peng2009, Bulbul2011}). However, these bounds depend on several assumptions and how observational data is interpreted, see e.g.~\citealt{Marin-Gilabert2024, Marin-Gilabert2025}. More recent and direct observations of the velocity dispersions in the Coma cluster could, on the contrary, be indicative of a large effective viscosity (\citet{XRISM:2025uvv}), although this also depends on the modeling and interpretation of the measurements (\citet{Vazza25}). 
Moreover, the suppression might differ substantially among clusters and it might also not be the same for different transport coefficients. 
Overall, the transport properties of the ICM are an open issue and they are subject of current research (see e.g.~\citet{Zhang2024}).   
Furthermore, it is important to note that direct helium abundance measurements that could shed light on this topic are simply not available. 
In order to estimate how a suppression of transport may affect our previous idealized results, we have solved the Burgers equations with the substitution $K_{\rm{p He}} \to K_{\rm{p He}}/f_B$, where $f_B$ is a suppression factor. In Fig.~\ref{fig:stellar_halo_mass_relation3} we show the ratio between stellar mass with and without sedimentation for $f_B=1/3$ and different values of $\alpha$. In the left plot we have taken $T = \tvir$, as in Figs.~\ref{fig:stellar_halo_mass_relation} and~\ref{fig:stellar_halo_mass_relation2}. It can be seen that helium sedimentation effects are very reduced with respect to the ideal case with $f_B=1$ and they give at most a factor $\sim 2$ correction for the largest clusters with $\mhalo \sim 5\times10^{15} \msun$.

On the other hand, the sedimentation process does show a strong dependence on the temperature, as $K_{\rm{pHe}} \propto T^{-3/2}$. To illustrate this, in the right plot of Fig.~\ref{fig:stellar_halo_mass_relation3} we have taken $f_B=1/3$ and a constant temperature 30\% larger, namely $T=1.3 \, \tvir$ (a conservative choice, see for instance \citet{nelson}). This partially compensates the suppression coming from $f_B$. More generally, it is important to note that realistic, radial-dependent temperature profiles, may yield larger effects than the idealized isothermal one we have considered. This is because in the isothermal scenario, although higher temperatures increase the velocity of sedimentation, they also give a more extended gas distribution, which tends to reduce the cooling rate in the inner regions of the cluster. A detailed study of helium sedimentation effects with more realistic temperature profiles lies outside the goal of this paper and is left for future work.     

The suppression of transport coefficients  has been associated to turbulence (see the above references for discussions and caveats), which in turn seems to be related to recent merger activity.  Clusters of galaxies show different levels of turbulence and merger activity, so it is possible that they could also present different sedimentation suppression factors.  For instance the galaxy cluster El Gordo, at $z\sim0.9$ with $M\sim2\times 10^{15} M_\odot$, is extremely massive and its BCG shows low star formation activity; morphological analysis of the gas emission suggests a recent merger while spectroscopic analysis shows that it has a high turbulence velocity of $\sim1000$ km$/$s (\citet{ElGordo}).  This might be indicative of a significant suppression factor of sedimentation and could be the reason for its low star formation rate. 
On the other hand, the Phoenix galaxy cluster, of similar mass at $z\sim0.6$ (\citet{Williamson2011}), has high star formation rate, grater than $ \sim 500 M_\odot/$yr (\citealt{McDonald2012, McDonald2015, Mittal2017}), and line broadening associated with a relatively low turbulence of $300$ km$/$s, and no evidence of a recent merger (\citet{Phoenix}).  It is tempting to think that the lack of recent merger activity and/or lower turbulence could be associated to a lower suppression factor in which case sedimentation would  favor star formation in this cluster.  This link has not been confirmed, but it lends support to the possibility of a wide range of suppression factors affecting different clusters of galaxies.

\begin{figure}[t]
\includegraphics[width=0.5\textwidth,{angle=0}]{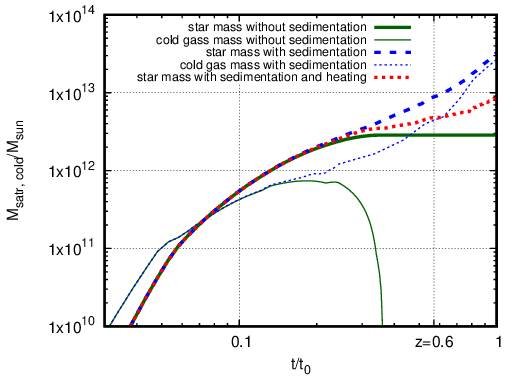}
\caption{\label{fig:phoenix_star_evolution} Stellar mass and cold gas mass (in units of solar mass) 
as a function of time (normalized to the present time $t_0$), for a galaxy cluster with $\mvir=5 \times 10^{15} M_{\rm sun}$ at the present time. The stellar masses without helium sedimentation, with helium sedimentation but without heating, and with helium sedimentation and heating are given by the thick green solid, blue dashed, and red dotted lines, respectively. In turn, the cold gas masses without helium sedimentation and with helium sedimentation (not including heating) are given by the thin green solid and blue dashed curves, respectively. As a reference, we have indicated with a vertical line the time corresponding to the redshift of the Phoenix cluster ($z \simeq 0.6$). For this plot we have taken $\alpha=0.05$.}
\end{figure}

\begin{figure*}[t]
\includegraphics[width=0.45\textwidth]{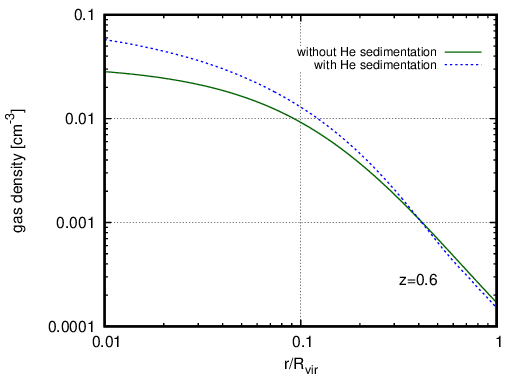} \quad
\includegraphics[width=0.45\textwidth,{angle=0}]{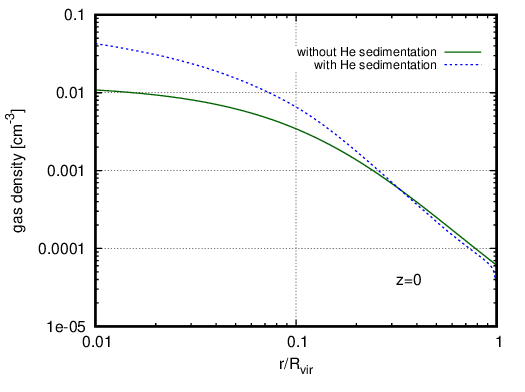}
\caption{\label{fig:gas_density} Gas density (in ${\rm cm}^{-3}$) as a function of radius (normalized to the virial radius at the corresponding redshift) for a cluster with $\mvir= 5 \times 10^{15} M_{\rm sun}$ at the present time. The green solid curve gives the density without sedimentation and the blue dashed curve gives the density with sedimentation. The left plot shows the density profiles at redshift $z=0.6$, when the cluster was younger and had a mass $\mvir \simeq 2 \times 10^{15} M_{\rm sun}$ (i.e., for a redshift and mass similar to the ones of the Phoenix cluster), while the right plot shows the density profiles at $z=0$.}
\end{figure*}

\begin{figure*}[t]
\includegraphics[width=0.45\textwidth]{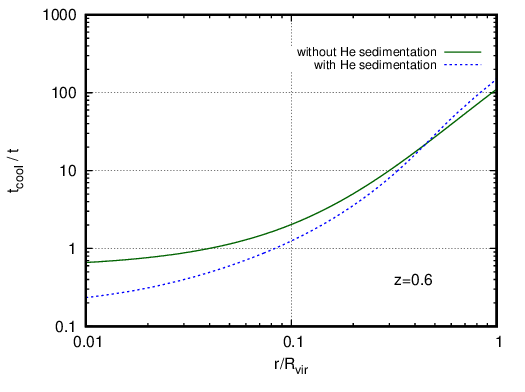} \quad
\includegraphics[width=0.45\textwidth,{angle=0}]{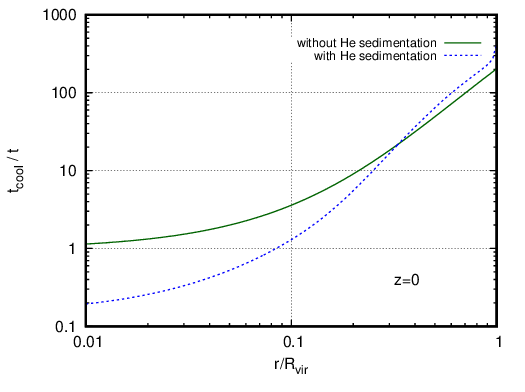}
\caption{\label{fig:cooling_time} Same as Fig.~\ref{fig:gas_density}, but plotting $\tcool/t$ instead of the gas density.}
\end{figure*}

One of the most interesting aspects of helium sedimentation effects is that, if present, they become important at late times. Therefore helium sedimentation provides a plausible physical mechanism for inducing a stage of late star formation. This can be seen in Fig.~\ref{fig:phoenix_star_evolution}, where we show the time evolution of the stellar and cold gas masses, for a galaxy cluster with $\mvir=5\times 10^{15} \, M_{\rm sun}$ at $z=0$. The star formation history is the same with or without sedimentation at times earlier than $\sim 0.3 \, t_0$. 
At later times, the simple model without helium sedimentation predicts basically no more star formation, while helium sedimentation leads to an era of renewed star formation or even to a starburst. 
Intriguingly, as mentioned above, a high starburst activity has been found in the central galaxy of the Phoenix cluster, which has a redshift $z \simeq 0.6$ (vertical line in the figure) and a mass $\sim 2\times 10^{15} \msun$, which, in the simplified accretion model we have considered, corresponds to a cluster with a mass $\mhalo \sim 5 \times 10^{15} \msun$ at $z=0$. As can be seen from Fig.~\ref{fig:phoenix_star_evolution}, helium sedimentation could lead to starburst activity at late times, and in particular at $z \sim 0.6$, as is the case of the Phoenix Cluster. 

To understand in more detail why this can happen, we compare in Fig.~\ref{fig:gas_density} the gas density profiles with and without helium sedimentation at redshift $z=0.6$, relevant for a tentative comparison with the Phoenix cluster (left plot) and also at $z=0$, when helium sedimentation has proceeded further (right plot). The total amount of gas inside $\rvir$ is the same with or without sedimentation, but it is substantially more concentrated in the center when there is sedimentation. Accordingly, the cooling time in the inner regions is much shorter, as can be seen in Fig.~\ref{fig:cooling_time}. In the isothermal model without sedimentation, the cooling time is larger than the age of the cluster for almost all radii at $z \sim 0.6$ and everywhere at $z=0$, therefore the amount of gas that can cool is negligible or essentially zero. Instead, when there is helium sedimentation, a significant amount of gas can cool and this amount increases with time, as seen in Fig.~\ref{fig:phoenix_star_evolution} (Figs.~\ref{fig:gas_density} and~\ref{fig:cooling_time} are two snapshots of the complete time evolution depicted in Fig.~\ref{fig:phoenix_star_evolution}).

\section{Conclusions and outlook}
\label{sec:conclusions}
Helium sedimentation has the potential to produce substantial changes in the gas density profile of galaxy clusters. Here we have studied how this can affect the history of star formation in central cluster galaxies. The main effect is that, as helium sediments, the gas density in the inner regions of the cluster increases, boosting the cooling rate and therefore the amount of cold gas that can form stars.
Helium sedimentation also results in a slow contraction of the gas (the center of mass moves inwards very slowly), and the gravitational energy that is released may, at least partially, heat the gas.

To obtain a quantitative estimate of these effects we have performed a 1-dimensional numerical simulation of helium sedimentation within a simple SAM of star formation without AGN feedback, as our current aim is to isolate the effects of helium sedimentation. In summary, our results show that helium sedimentation effects on the star formation rate are irrelevant in clusters with a mass $\mhalo \lesssim 10^{14} \msun$, the effects seem at most mild in clusters with $10^{14} \msun \lesssim \mhalo \lesssim 10^{15} \msun$, and they can potentially be very significant in the largest clusters, $\mhalo \gtrsim 10^{15} \msun$ (see Figs.~\ref{fig:stellar_halo_mass_relation} and~\ref{fig:stellar_halo_mass_relation2}).
The strong dependence on the mass arises mainly because the velocity of helium sedimentation scales with the temperature to the power of 3/2. We have also found that the gravitational heating due to helium sedimentation can be important, partially countering the increase in the cooling rate. Hence, it could be interesting to develop a more detailed treatment of this heating effect.

Sedimentation can be hindered by magnetic fields and plasma instabilities, which can be represented, as a first approximation, by a suppression factor. Our calculations indicate that a suppression factor of 1/3 in the diffusion velocity of helium already reduces the effects considerably (see Fig.~\ref{fig:stellar_halo_mass_relation3}), while a suppression factor around 1/5 seems enough to inhibit any effects of helium sedimentation on star formation. However, it is important to stress that the magnitudes of the effects we have obtained correspond to an isothermal model, while more realistic, radially dependent temperature profiles, could substantially increase the effects of sedimentation and partially compensate suppression contributions (this further analysis is left for future work). It is also worth noticing that clusters in different dynamical states, or with different recent merger histories, appear to show wildly different turbulence velocities, which could indicate that the suppression can take an extended range of values.

An interesting and distinctive issue is that, since helium sedimentation is a very slow process, the effects (if any) arise at late times (see Fig.~\ref{fig:phoenix_star_evolution}). Therefore helium sedimentation is a possible physical process that can drive an era of late star formation in a cluster (as e.g. observed in the Phoenix cluster).

\begin{acknowledgments}
J. R. thanks the support by a grant Consolidar-2023-2027 from the Secretar\'ia de Ciencia y Tecnolog\'ia \mbox{(SeCyT)} de la Universidad Nacional de C\'ordoba (UNC). J. R. and N. P. acknowledge support by a grant PIP Raices Federal 29320230100004CO from the Consejo Nacional de Investigaciones Cient\'ificas y T\'ecnicas \mbox{(CONICET)}, Argentina.  N. P. acknowledges support from PICT 2021-00700 and PICT-Raices Federal 2023-0002.
\end{acknowledgments}

\bibliography{he_sed_star_form}

\end{document}